\documentclass[twocolumn,secnumarabic,amssymb, nobibnotes, aps, prd, superscriptaddress]{revtex4-1}
\usepackage[T1]{fontenc}
\usepackage{graphicx}
\usepackage{amsmath}
\usepackage{upgreek}
\usepackage{SIunits}

\begin{document}
\title{Comparison of the slip of a PDMS melt on weakly adsorbing surfaces measured by a new photobleaching-based technique}

\author{Marceau H\'enot}
\affiliation{Laboratoire de Physique des Solides, CNRS, Univ. Paris-Sud, Universit\'e
Paris-Saclay, 91405 Orsay Cedex, France}
\author{Alexis Chennevi\`ere}
\affiliation{Laboratoire de Physique des Solides, CNRS, Univ. Paris-Sud, Universit\'e
Paris-Saclay, 91405 Orsay Cedex, France}
\affiliation{Laboratoire L\'eon Brillouin CEA, CNRS, CEA Saclay 91191, Gif sur Yvette Cedex, France}
\author{Eric Drockenmuller}
\affiliation{Univ Lyon, Universit\'e Lyon 1, CNRS, Ing\'enierie des Mat\'eriaux Polym\`eres, UMR 5223, F-69003, Lyon, France}
\author{Liliane L\'eger}
\affiliation{Laboratoire de Physique des Solides, CNRS, Univ. Paris-Sud, Universit\'e
Paris-Saclay, 91405 Orsay Cedex, France}
\author{Fr\'ed\'eric Restagno}
\email[Corresponding author: ]{frederic.restagno@u-psud.fr}
\affiliation{Laboratoire de Physique des Solides, CNRS, Univ. Paris-Sud, Universit\'e
Paris-Saclay, 91405 Orsay Cedex, France}
\date{\today}

\begin{abstract}


We present an experimental method allowing to quantify slip at the wall in viscous polymer fluids, based on the observation of the evolution under simple shear flow of a photobleached pattern within a fluorescent labeled polymer melt. This straightforward method provides access to slip length at top and bottom interfaces in the $1~\upmu$m to 1~mm range and to the actual shear rate experienced by the fluid. Based on simple optical imaging and image analysis techniques, this method affords an improvement compared to previously reported methods in which the photobleached fluorescence intensity profiles before and after shear were compared and measured by scanning a photomultiplier. The present method relies on a direct determination of the displacement profile inside the polymer fluid from an analysis of the 3D evolution of the whole photobleached pattern. We demonstrate the potential of this method with measurements of the slip length for an entangled PDMS melt, as a function of the shear rate, in contact with several weakly surfaces i.e. end-tethered PDMS or polystyrene (PS) chains, a self-assembled monolayer (SAM) of trimethoxy(octadecyl)silane (OTS), and a glassy PS thin-film.
\end{abstract}
\maketitle


\section{Introduction} 
When a fluid flows between solid walls, the question of the boundary condition for the tangential velocity at fluid/solid interfaces is never an easy one to deal with~\cite{navier,churaev1984slippage,ramamurthy_1986}. The simplest approximation, often experimentally valid, is to consider that the relative fluid/wall tangential velocity is equal to zero. This approximation has however been questioned in recent years, especially because of the emergence of experimental techniques able to test, often indirectly, this approximation~\cite{neto2005boundary,lauga_2007}. In the case of complex fluids such as polymer fluids, it has been shown that this assumption is not always correct, and slip at interfaces has been early invoked to explain the origin of the flow instabilities during extrusion of polymer melts~\cite{rielly_1961}. This phenomenon has been widely investigated in recent years, based on experiments trying to access more directly the local flow velocity near the solid wall~\cite{churaev1984slippage,migler_slip_1993,durliat_influence_1997,cottin-bizonne_nanorheology:_2002,craig_shear-dependent_2001,neto2005boundary}. Because it is quite difficult to measure the flow velocity at distances from the solid wall comparable to molecular sizes, the physical quantity that is usually used to characterize slip at the wall is the slip length $b$, defined as the distance from the wall where the velocity profile linearly extrapolates to zero. This quantity has been reported to range from few to hundreds of nanometers for simple fluids or even to hundreds of micrometers for complex fluids~\cite{pit_2000,schmatko_2005,leger_wall_1997,durliat_influence_1997,massey_investigation_1998, neto2005boundary,bocquet2007flow,lauga_2007,cottin-bizonne_2008}. The question of slip at interfaces for polymer fluids, either melts or entangled solutions, has been actively studied theoretically and experimentally for the last 20 years~\cite{de_gennes_1979,hatzikiriakos2012wall}, both for its conceptual and practical importance in many fields such as polymer assisted oil recovery~\cite{de1989slip,de1989slip,chauveteau1981basic,cuenca_fluorescence_2012,cuenca_submicron_2013}, polymer extrusion~\cite{ramamurthy_1986,piau1994measurement,denn2001extrusion}, lubrication in industrial processes~\cite{cayer2008drainage} or in biological systems~\cite{dedinaite_2012}. Indeed, measuring a slip velocity or a slip length can give direct information on the friction mechanisms between the liquid and the surface, and can thus lead to controlled manipulations of the friction through specifically designed surface modifications~\cite{migler_slip_1993, mcgraw_nanofluidics_2014}. A number of experimental techniques have been developed in order to probe the slip behavior of polymer fluids on various surfaces. Most are indirect techniques, but only a few of them lead to direct measurements of the velocity close to the solid wall. These techniques include the measurement of the dewetting kinetics of thin polymer films~\cite{brochard_1996,reiter_real-time_2000,baumchen_slippage_2012,baumchen_reduced_2009}, the observation of the evolutions of microstructures pre-made in glassy polymers when heated above the glass transition temperature using atomic force microscopy~\cite{mcgraw_nanofluidics_2014, mcgraw_slip-mediated_2015}, the studies of Plateau-Rayleigh instabilities on a fiber~\cite{haefner_influence_2015}, the direct rheological measurement of the force required to drain polymer fluid between surfaces~\cite{neto2005boundary}, neutron reflectometry~\cite{chenneviere_direct_2016}, or the combination of rheometry and confocal fluorescence microscopy~\cite{horn2000hydrodynamic, boukany_molecular_2010}. Thus, it must be pointed that only a few techniques allow the direct measurement of the solid-liquid velocity~\cite{migler_slip_1993,lumma2003flow}.

The velocimetry technique based on the observation of the evolution of a pattern drawn in a fluorescent fluid using photobleaching has proved its efficiency for fluid-solid velocity measurement~\cite{migler_slip_1993,durliat_influence_1997,leger_wall_1997,massey_investigation_1998,cuenca_fluorescence_2012,cuenca_submicron_2013,schembri_velocimetry_2015}. The evolution of this motif when the fluid flows enable access to the slip velocity and the slip length. The motif can be an interference pattern drawn in the fluid near a wall using evanescent waves~\cite{migler_slip_1993,leger_wall_1997}: in that case, the average velocity inside the penetration depth of the evanescent wave can be deduced directly from the period of oscillation of the fluorescence intensity when the photobleached pattern is displaced by the flow in front of the reading interference pattern of light. With penetration depth in the range of 50~nm, this is close to the velocity of the first molecular layer at the solid wall, for polymers of molecular weight higher than a few hundreds of kg$\cdot$mol$^{-1}$. The pattern can also be a simple line going throughout the fluid~\cite{leger_wall_1997, cuenca_fluorescence_2012,cuenca_submicron_2013,schembri_velocimetry_2015}. The evolution of this line under the effect of the flow, in particular the displacements at both surfaces and the inclination due to the shear rate, contains all the information on the flow characteristics.

The aim of the present article is to present a significantly improved version of this printed line technique. In a first part, we describe the experimental setup and the typical polymer/solid interfaces probed in the present work in order to demonstrate the potential of this new technique. We then show in details how the pioneering experiment of Durliat, Massey, Migler~\textit{et al.}~\cite{migler_slip_1993,durliat_influence_1997,massey_investigation_1998} can be improved by replacing the scan of the photobleached pattern using a photomultiplier tube as light sensor by an image acquisition setup and a detailed image analysis. In a third part, we show that this imaging technique can be further improved in terms of accuracy in the determination of slip at the wall, based on the possibility of achieving local intensity measurements on the image, leading to a spatial resolution of the local velocity as a function of the depth inside the fluid, in the direction normal to the surfaces. We compare the two imaging methods, discuss their respective limitations and accuracy, and we examine the effect of diffusion. Finally, the potential of these imaging methods is demonstrated through measurements of the slip velocities for a high molecular weight entangled polydimethylsiloxane (PDMS) melt flowing on various weakly adsorbing surfaces.

\section{Material and methods}
\begin{figure}[htbp]
  \centering
  \includegraphics[width=240pt]{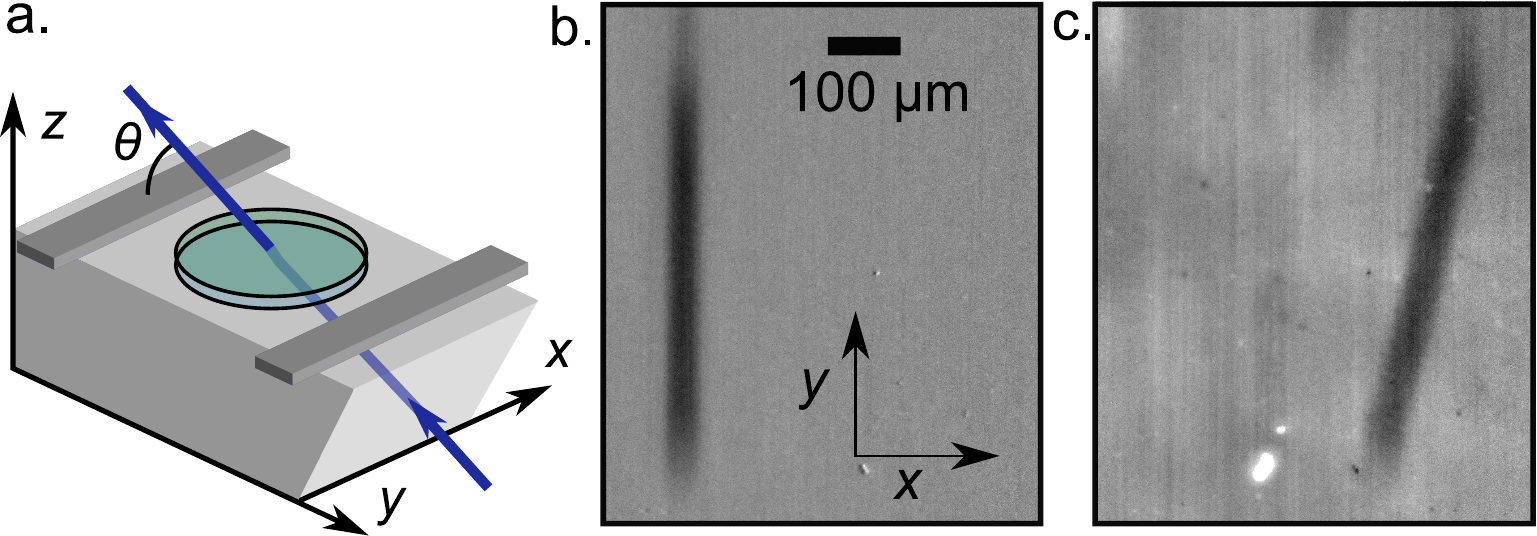}
  \caption{(a) Schematic of the experimental setup: a drop of liquid is deposited on a silica prism. The mobile top plate sitting on two spacers is not represented. Images of the fluorescence of the sample, with a well visible photobleached line, before (b) and after (c) shearing are shown. The total displacement of the top plate, in the $x$ direction, is $d_\mathrm{s} = 408\pm 5~\upmu$m.}
  \label{schema_FRAP}
\end{figure}
{\bf Experimental setup.} The setup is presented in Figure~\ref{schema_FRAP}a. A drop of a fluorescent photobleachable liquid is sandwiched between a silica prism at the bottom and a silica plate at the top. They are separated by $100\pm 2~\upmu$m thick polyester spacers. The liquid forms a disk of typical diameter 10-15~mm. Care is taken to ensure that the liquid does not touch the spacers. The observation of an interference pattern due to a neon light illumination allows the adjustment of the parallelism of the two facing surfaces. The exact thickness $h$ is measured using spectroscopic reflectometry with normal white light illumination (ocean optics USB4000). The sample is illuminated by a laser at $\lambda= 458$~nm (Innova 90C) going from the bottom prism to the top plate, in the $(y,z)$ plane with an angle $\theta$ with respect to $y$. A lab constructed microscope made of a long working distance 10x objective (Mitutoyo) and a tube lens (Mitutoyo MT-1) images the fluorescence of the sample from the top, using a Nikon camera (D3200). A longpass filter at 500~nm (Edmund) is placed between the objective and the tube lens to stop the direct signal from the laser.
Two modes of illumination of the sample by the laser beam are available. The reading mode corresponds to a beam diameter of 2~mm when it reaches the sample. In this mode, the fluorescence can be imaged on an area of $1.5\times 2.2$~mm. The writing mode is obtained by placing a lens of focal length 10~cm just before the prism, which focuses the beam into the liquid. In this mode, the beam diameter is around $20-30~\upmu$m inside the liquid. This focused beam is used to photobleach a line in the fluorescent liquid by illuminating it during 800~ms with a 20~mW incident power. Switching between the two modes is achieved by displacing the lens with a motorized wheel (Thorlabs FW102C).
The fluid can be sheared in the $x$ direction by moving the top plate at a chosen constant velocity over a distance $d_\mathrm{s}$ during a time $t_\mathrm{s}$. This displacement is actuated by a stepper motor and is monitored directly using a LVDT sensor. Images of the photobleached pattern before and after shear are presented in Figures~\ref{schema_FRAP}b and~\ref{schema_FRAP}c. The bottom prism can be replaced with a 3~mm thick silicon wafer in which case the illumination comes from the top silica plate.

{\bf Polymer fluid and surfaces.} The polymer melt used in the present experiments is a trimethylsiloxy terminated PDMS melt with a numberaverage molecular weight $M_\mathrm{n}=608\times 10^{3}$~g$\cdot$mol$^{-1}$ and a chain dispersity \DJ~$=1.15$ (fractionated from a commercial batch, Petrarch PS047.5) containing 0.5~w$\%$ of lab-made fluorenscently-labelled PDMS ($M_\mathrm{n}=320\times 10^{3}$~g$\cdot$mol$^{-1}$,  \DJ~$=1.18$ labelled at both ends with nitrobenzoxadiazole groups) emitting at 550~nm~\cite{leger1996,cohen_synthesis_2012}.
The top surface was made of fused silica and was cleaned with a piranha solution~\cite{kern_1970} just before assembling the flow cell. Different bottom surfaces were used in this study. A grafted layer of PDMS was made on a silica prism by grafting short PDMS chains with $M_\mathrm{n}=2\times 10^{3}$~g$\cdot$mol$^{-1}$ by the "grafting-to" technique, with a dry thickness of $z^*=3.2$~nm corresponding to the onset of the stretched brush regime. The advancing contact angle of water of this surface was $\theta_\mathrm{a}=112^\circ$ with an hysteresis of $5^\circ$. The grafting procedure and the details of the synthesis of the grafted chains are reported in L\'eger \textit{et al.}~\cite{Leger1999}. The OTS surface was made following a procedure detailed by Pit~\textit{et al.}~\cite{pit_2000} by immersing a silicon wafer previously cleaned by UV/Ozone in a bath of anhydrous hexadecane and CCl$_4$ (0.9:0.1 $V/V$), freshly distillated OTS (Sigma, 95$\%$) and a water saturated mixture of CCl$_4$ and chloroform (0.6:0.4 $V/V$). The bath temperature was 18$^\circ$~C. The advancing contact angle of dodecane of this surface was $\theta_\mathrm{a}=34^\circ$ with an hysteresis of $1^\circ$. A grafted layer of PS ($M_\mathrm{n}=350\times 10^{3}$~g$\cdot$mol$^{-1}$,  \DJ~$=1.25$, Polymer Source) was made on a silicon wafer using the procedure detailed in ref.~\cite{henot_influence_2017} leading to a dry thickness of 5.8~nm. The PS film was made by spincoating a 3~wt\% solution of PS ($M_\mathrm{n}=422\times 10^{3}$~g$\cdot$mol$^{-1}$,  \DJ~$=1.05$) in toluene at 2000 rpm. Its thickness after 24~h of annealing at 140$^\circ$~C under vacuum was 105~nm.

The two methods for slip length determination are both based on an analysis of the deformation of a line obliquely photobleached into the fluid, due to the shear flow. The first method, which is the only one available for very thin samples, globally analyses the z-integrated fluorescence intensity. The second one, best adapted to thicker samples, benefits of the fact that the bleached line is printed at a small angle with respect to the $y$ axis, so that analyzing the evolution of the photobleached line locally along the $y$ axis gives access to vertical resolution for the displacement field inside the fluid, and thus to a full determination of slip lengths at both the top and bottom surfaces, along with a measurement of the shear. The early method described in the literature was based on a comparison between the fluorescence intensity profiles of a bleached line before and after shear obtained through mechanical scans of the photomultiplier tube. It is worth pointing out that in the imaging techniques we propose here, there is no need of any mechanical scan, meaning that all uncertainties  associated to the mechanical backlash of the translation stages are avoided in the determination of the displacements inside the fluid. This will be emphasized later in the article.

\section{Slip Measurements using a z-integrated fluorescence intensity profile}
\begin{figure}[htbp]
  \centering
  \includegraphics[width=240pt]{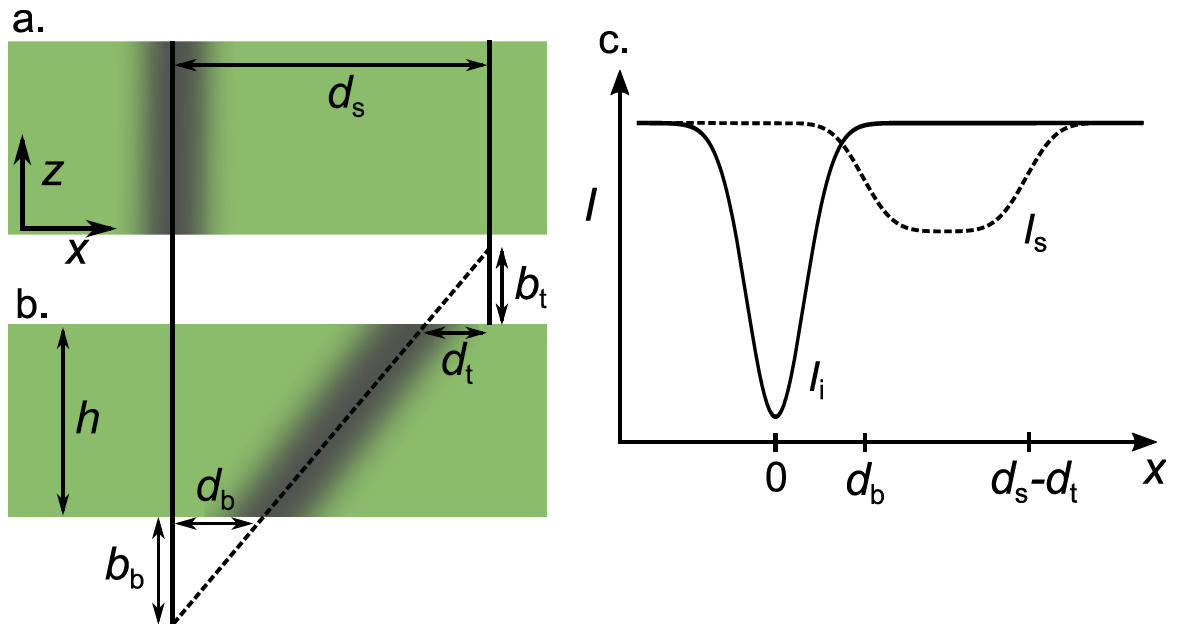}
  \caption{Principle of velocimetry using z-integrated intensity analysis of a photobleached line in a fluorescent polymer fluid. In (a), the theoretical photobleached line is represented in the fluid. Its intensity profile, integrated over the $z$ axis, $I_\mathrm{i}$ appears in (c) in plain line. In (b), the liquid has been sheared over a distance $d_\mathrm{s}$, it has slipped on the top and bottom plates over distances $d_\mathrm{t}$ and $d_\mathrm{b}$ respectively. The corresponding intensity profile $I_\mathrm{s}$ is drawn as the dash line in (c).}
  \label{principe_FRAP}
\end{figure}

{\bf General principle.} For low Reynolds number flows, the displacement of the top plate induces a Couette flow with a linear displacement profile $d(z)$ inside the fluid. As discussed later, the diffusion of polymer chains in the fluid can be neglected in a first approach. Because of wall slip, the displacement profile depends on the slip lengths as can be seen in Figures~\ref{principe_FRAP}a and~\ref{principe_FRAP}b.

{\bf Intensity profiles.} Due to the small thickness of the flow cell, the fluorescence intensity profile can only be monitored from above. Theoretically, the intensity profile over the $x$ direction (cf Figure~\ref{principe_FRAP}c) in the mid-plane of the bleached line allows the determination of the slip distances both at the top and bottom interfaces $d_\mathrm{t}$ and $d_\mathrm{b}$, and thus, the shear rate experienced by the fluid: $\dot{\gamma}=(d_\mathrm{s}-d_\mathrm{t}-d_\mathrm{b})/(t_\mathrm{s} h)$. The slip lengths at the top and bottom interfaces, $b_\mathrm{t}$ and $b_\mathrm{b}$, can thus be extracted using the following relation:
\begin{equation}
b_\mathrm{t,s}=\frac{d_\mathrm{t,s}h}{d_\mathrm{s}-d_\mathrm{t}-d_\mathrm{b}}
\label{equ_bd}
\end{equation}

The initial fluorescence intensity profile is denoted $I_\mathrm{i}(x)$. If the energy used to bleach the fluorescence is kept low enough, this initial profile is well approximated by a Gaussian function of width $\sigma_0$:
\begin{equation}
I_\mathrm{i}^\mathrm{G}(x)=I_0-A\exp \left(-\frac{x^{2}}{\sqrt{2}\sigma_0}\right)
\label{eq_Ii}
\end{equation}
After shear, the profile $I_\mathrm{s}(x)$ is given by:
\begin{equation}
I_\mathrm{s}(x)=\frac{1}{h}\int_0^{h} I_\mathrm{i}(x-d(z))\mathrm{d}z
\label{eq_int_cis}
\end{equation}
\begin{equation}
I_\mathrm{s}^\mathrm{G}(x)=I_0-B\left[\mathrm{erf}\left(\frac{x-d_\mathrm{b}}{\sqrt{2}\sigma_0}\right)
-\mathrm{erf}\left(\frac{x-d_\mathrm{s}+d_\mathrm{t}}{\sqrt{2}\sigma_0}\right)\right]
\label{eq_Is}
\end{equation} with:
\begin{align*}
B=\frac{A\sigma_0\sqrt{\pi}}{2(d_\mathrm{s}-d_\mathrm{t}-d_\mathrm{b})}
\end{align*}

\begin{figure}[htbp]
  \centering
  \includegraphics[width=160pt]{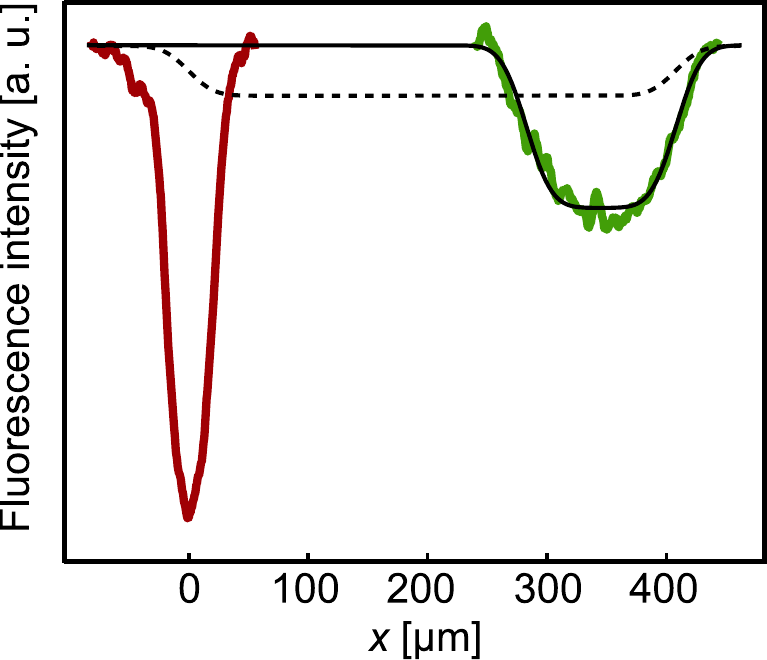}
  \caption{The fluorescence intensity profiles $I_\mathrm{i}(x)$ and $I_\mathrm{s}(x)$ determined from the integration over $y$ of the images shown in Figure~\ref{schema_FRAP} are represented respectively in red and green. The solid black curve is the best profile calculated from equation~\ref{eq_int_cis} and the initial profile. This gives $b_\mathrm{t} = 1 \pm 5~\upmu$m and $b_\mathrm{b} = 225\pm 20~\upmu$m. The dashed curve corresponds to the case of no slip.}
  \label{profils_exp}
\end{figure}

{\bf Experimental results and analysis.} 
In the first method, the z-integrated fluorescence intensity profiles can be extracted from the images by integrating the signal over the $y$ direction. Such integrated intensity profiles $I_\mathrm{i}(x)$ and $I_\mathrm{s}(x)$, before and after shear respectively, deduced from the images of Figure~\ref{schema_FRAP}, are shown in Figure~\ref{profils_exp}. The sheared profile has been rescaled so that its area matches that of the initial profile. The Gaussian function $I_\mathrm{i}^\mathrm{G}$ is fitted on the initial profile in order to determine the parameters $I_0$, $A$ and $\sigma_0$. The distance $d_\mathrm{s}$ is measured directly, following dusts on the images. Using these parameters, the function $I_\mathrm{s}^\mathrm{G}$ is fitted to the sheared profile with only the two slip distances as adjustable parameters. It is checked that the profile numerically calculated from the initial profile and equation~\ref{eq_int_cis} matches the sheared profile as shown in Figure~\ref{profils_exp}.

{\bf Precision of the method.}
The situation allowing the best accuracy on $b_\mathrm{b}$ is the case of very low slip at the top interface: $b_\mathrm{t} \ll h$. In this case, $b_\mathrm{b}/h$ is given by:
\begin{equation}
\frac{b_\mathrm{b}}{h}=\frac{1}{d_\mathrm{s}/d_\mathrm{b}-1}
\end{equation}
Therefore, the relative uncertainty on $b_\mathrm{b}/h$: $\frac{\delta(b_\mathrm{b}/h)}{b_\mathrm{b}/h}$ is minimum when $b_\mathrm{b}=h$, reaching $3~\%$ for an uncertainty on $d_\mathrm{b}$  $\delta(d_\mathrm{b})=2~\upmu$m. This uncertainty increases strongly when $d_\mathrm{b}$ comes close to $d_\mathrm{s}$. In order to stay below $20~\%$, $b_\mathrm{b}/h$ has to be less than 20. There is no real upper experimental limit in $h$, hence very large slip can be measured accurately. On the other hand, low values of $h$ are required in order to measure weak slip on the bottom surface. The minimum thickness of spacers that we used is 10~$\upmu$m. In this case, for low slip, $d_\mathrm{s}$ is limited because it is inversely related to the average intensity of the profile, hence to the signal to noise ratio. For $d_\mathrm{s}=200~\upmu$m, the minimum value of $b_\mathrm{b}$ that can be measured with a relative uncertainty less than $20~\%$ is 0.5~$\upmu$m. The uncertainty on the slip length is larger at the top surface than at the bottom one because it is directly related to that on $d_\mathrm{s}$, $\delta(d_\mathrm{s})= 5~\upmu$m.

{\bf Effect of diffusion.}
The diffusion of the labeled molecules in the fluid has not been taken into account in the above analysis of the data. The diffusion causes a broadening of the profiles that could be mistaken and interpreted as a reduction of slip. Because the fluorescence signals are integrated over $y$ and $z$, there is here no effect of the amplification of diffusion due to shear~\cite{taylor_1954, cuenca_fluorescence_2012}. The typical root mean square width of the initial profile is $\sigma_0 = 15~\upmu$m. Therefore the diffusion has an effect on the uncertainty of $b$ when $D$ is higher than $D_\mathrm{max} = [(\sigma_0+\delta(d_\mathrm{b}))^2-\sigma_0^2]/(\tau + t_\mathrm{s})$ where $\tau = 10$~s is the time needed to acquire the images. The longest shear duration used in this study was 50~s leading to $D_\mathrm{max} = 10^{-12}$~m$^{2}\cdot$s$^{-1}$. The diffusion coefficient of the fluorescent polymer chains in the PDMS melt used here~\cite{leger1996} is below $10^{-15}$~m$^{2}\cdot$s$^{-1} \ll D_\mathrm{max}$ which justifies that diffusion can indeed be neglected. The minimum shear rate that could be applied corresponds to $t_\mathrm{s}\approx 1000$~min leading to $\dot{\gamma}_\mathrm{min}=2\times 10^{-5}$~s$^{-1}$ for $d_\mathrm{s}-d_\mathrm{t}-d_\mathrm{b} = 100~\upmu$m.

\section{Measure of the slip and of the full displacement profile using a z-resolved method}
\begin{figure}[htbp]
  \centering
  \includegraphics[width=250pt]{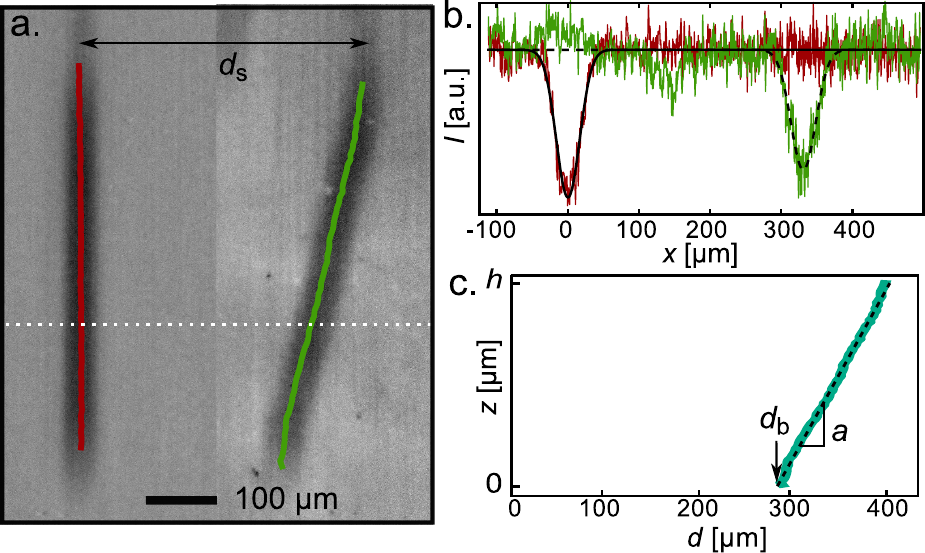}
  \caption{Direct determination of the characteristics of the velocity profile inside the fluid, obtained by measuring the local displacement of slices of the pattern in the images before and after shearing as a function of the position $y$. In (a) half of the images of Figure~\ref{schema_FRAP} are reproduced, before (left) and after (right) shear. The displacements are obtained by fitting gaussian functions of the local (in $y$) fluorescence intensity profiles. Initial and sheared profiles corresponding to the white dashed line in (a) are shown in (b) in red and green respectively. The resulting displacement field $d(z)$ is shown in (c) and leads to $b_\mathrm{t} = 1 \pm 5~\upmu$m and $b_\mathrm{b}/h = 2.39 \pm 0.03$.} \label{profil_vitesse}
\end{figure}

We show now that the imaging setup can allow a full determination of the displacement field inside the fluid, after shear. The photobleaching laser beam goes through the liquid with an angle $\theta$ close to $10^\circ$ with respect to the $y$ axis. Hence, following on the images the fluorescence intensity profiles for different positions in the $y$ direction, it is possible to determine the displacement field $d(z)$. The angle $\theta$ can be measured accurately by analyzing in the initial image (Figure~\ref{schema_FRAP} b) the length of the pattern along the $y$ direction. Then, the local displacement of the fluid can accurately be determined, as shown in Figures~\ref{profil_vitesse}a and~\ref{profil_vitesse}b, by adjusting gaussian functions for various slices of the pattern at varying $y$ values. A typical displacement field thus determined is shown in Figure~\ref{profil_vitesse}c. A linear regression on $d(z)$ gives $b_\mathrm{b}$ and $b_\mathrm{t}$.

{\bf Precision of the method.} This method of slip length measurements is less impacted by the presence of dust on the surfaces than the z-integrated one because it does not involve integration of the images. Hence, the total displacement $d$ is only limited by the field of the camera. From a linear regression on the displacement profile, it is possible to extract $d(0)=d_{b}$ and the slope $a$ (cf Figure~\ref{profil_vitesse}c). The uncertainties on these parameters are typically $\delta(d_{b}) = 0.5~\upmu$m and $\delta(a) = 1\times 10^{-3}$. For low values of $d$, the uncertainty on $d_{b}$ causes $\delta(b_\mathrm{b}/h)$ to rise. It stays bellow $0.02$ when $d<300~\upmu$m and reaches $0.01$ for $d_\mathrm{s}=600 - 700~\upmu$m. Contrary to the integrated method, it is not necessary to know the exact values of $d_\mathrm{s}$ to measure the slip length at the bottom $b_\mathrm{b}$, which improves the accuracy. It should be noted that in the present work, the accuracy on the bottom slip length is limited by the uncertainty on $h$, $\delta(h) = 3~\upmu$m.  A limitation of this z-resolved method compared to the z-integrated one is the fact that it requires a gap $h$ between the plates at least several times higher than $\sigma_0$ while in the first method $h$ can easily be decreased down to $10~\upmu$m. Therefore the maximum shear rate $\dot{\gamma}_\mathrm{max~2}=20$~s$^{-1}$ that can be imposed in case of no-slip using this method is lower than in the z-integrated method $\dot{\gamma}_\mathrm{max~1}=200$~s$^{-1}$. 

{\bf Effect of diffusion.} Here also the diffusion of the fluorescent markers can affect the slip measurement. Indeed the diffusion along $z$ before and during shear tends to modify the shape of the motif, which, for large enough diffusion, should depart from a straight line in the image of the sheared sample. The criterion on $D$ remains similar to that for the z-integrated method, and is well verified in the present case. It is possible to check that the diffusion has no effect by verifying in Figure~\ref{profil_vitesse}b that there is no significant difference between the width of the fluorescence intensity profiles before and after shear.

\section{Application to the slip length measurement of an entangled PDMS melt on weakly adsorbing surfaces} 

\begin{figure}[htbp]
  \centering
  \includegraphics[width=240pt]{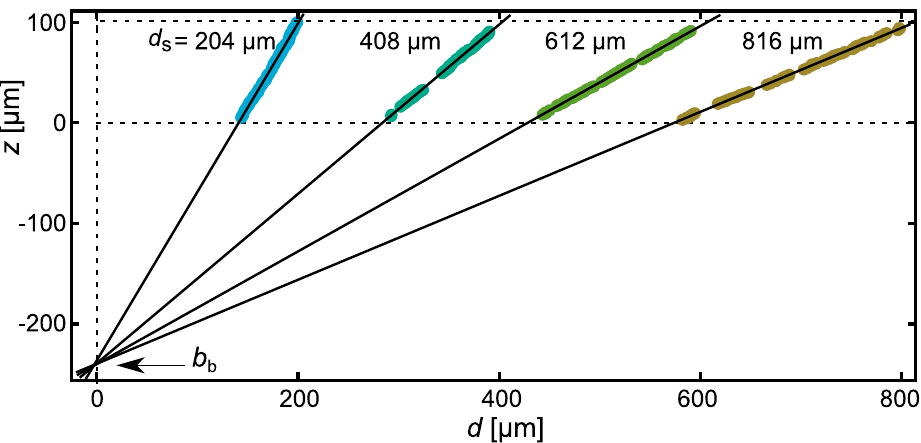}
  \caption{Displacement fields inside the fluid slipping on a grafted layer of PDMS for different total displacements $d_\mathrm{s} = 204$, $408$, $612$, $816~\upmu$m of the top plate. All these displacement profiles give $b_\mathrm{b}/h = 2.39 \pm 0.03$, and thus a bottom slip length in $b_\mathrm{b} = 239 \pm 7~\upmu$m.} 
  \label{profils_vitesse_d}
\end{figure}
Both methods described above have been used to measure the slip behavior of an entangled PDMS melt ($M_\mathrm{n}=608$~kg\cdot mol$^{-1}$) on short grafted PDMS chains ($M_\mathrm{n}=2$~kg\cdot mol$^{-1}$) that can prevent the adsorption of the melt chains onto the surface. In order to maximize the accuracy, the slip was measured at the bottom surface. An adsorbing surface was chosen as top surface in order to strongly reduce slip on that surface~\cite{massey_investigation_1998}. We could indeed verify that the top slip length was lower than the sensitivity of the setup. The velocity of the top plate during shear was $V_\mathrm{s}=1300~\upmu$m$\cdot$s$^{-1}$ and the gap between the two solid surfaces was $h=100\pm 3~\upmu$m. The z-integrated method gave $b_\mathrm{b} = 225 \pm 20~\upmu$m while the z-resolved method gave $b_\mathrm{b} = 239 \pm 7~\upmu$m as shown in Figure~\ref{profils_vitesse_d}. 

\begin{figure}[htbp]
  \centering
  \includegraphics[width=230pt]{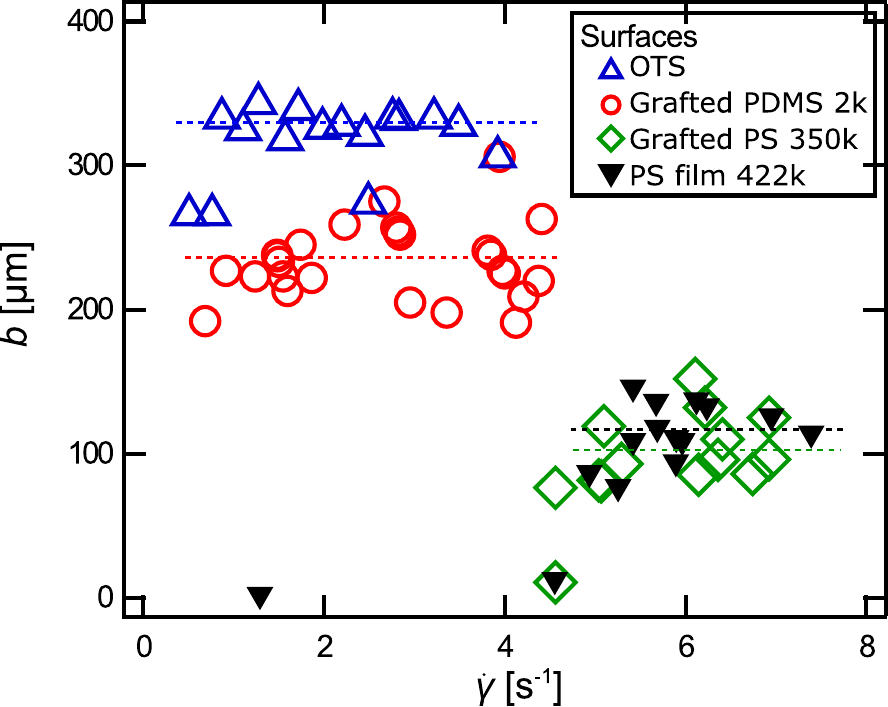}
  \caption{Slip length of a 610~kg$\cdot$mol$^{-1}$ PDMS melt on different weakly adsorbing surfaces as a function of shear rate. Horizontal dashed line correspond to average values on the constant regime.} 
  \label{glissement_surfaces}
\end{figure}

Using the z-resolved method, measurements were conducted on 4 different surfaces made of OTS, PS film, grafted layers of PS or PDMS and for different values of the top plate velocity. The results are shown in Figure~\ref{glissement_surfaces} as a function of the shear rate experienced by the fluid.  All these surfaces are expected to be weakly adsorbing.
In all cases, a regime of constant slip appears for the larger shear rates. The dispersion of the measurements seems to come from the inhomogeneity of the surfaces rather than the uncertainty of the method. Indeed the lowest dispersion is achieved on the measurements performed on the OTS surface despite the fact that the ratio $b/h$ is the least favorable. For the lower shear rates, we observe a decrease of the slip length (this is more evident in the PS case). This is reminiscent of the three regimes of slip previously observed experimentally~\citep{migler_slip_1993, massey_investigation_1998} and discussed theoretically~\citep{brochard_1996} in the literature: at low shear rate the slip length is lower than the method sensitivity, then there is a regime of non-linear friction where the slip length raises at almost constant shear rate followed by a regime of ideal slip with constant slip length. In the present measurements we concentrated our efforts on the regime of large constant slip that we measured with a high accuracy.

These three successive slip regimes when increasing the shear rate has been considered as the signature of the presence of adsorbed chains on the surfaces that entangled with the melt~\cite{massey_investigation_1998}. One open question is the mechanism of adsorption on this non-adorbing surfaces. A possible mechanism could be due to the presence of holes in the layers as the threshold is usually strongly dependent of the quality of the surfaces. However the fact that the threshold of the high slip regime is the same for the 5~nm thick grafted layer of PS and for the 105~nm thick PS film forbid in this case to attribute this effect to holes or to the incomplete screening of the underlying silicon wafer. A small adsorption is thus possibly measured by the onset of low friction. We did not achieved to measure any adsorption on the PS films using ellipsometry (resolution 0.5 nm).

It also appears from these measurements that the friction associated with the slip of a melt on a surface is strongly dependent of the interaction between the two. 
In these experiment, the slip length $b$ can be related to the friction stress at the wall:
\begin{equation}
\sigma=\eta\left(\frac{\partial v}{\partial z}\right)_{z=0}=\eta \frac{v(0)}{b}=kv(0)
\end{equation}
where $v(0)$ is the melt velocity close to the wall and $k$ a constant friction coefficient. This leads to $b=\eta/k$. The slip length is thus the ratio of a bulk property (the viscosity) and an interface property ($k$). De Gennes, proposed that $k$ should be independent of the entanglement and should be rather a measurement of the monomer-solid friction \cite{de_gennes_1979}. In our measurements, the average slip length in the constant regime goes from 110$~\upmu$m on PS to 235$~\upmu$m on PDMS grafted layer and 330$~\upmu$m on a SAM of OTS. These slip lengths are consistent with what was observed for PS by McGraw~\textit{et al.} on various surfaces~\cite{mcgraw_nanofluidics_2014}. As for simple fluids, it thus appear that the friction is really sensitive to the chemical structure of both the fluids and the solids \cite{neto_boundary_2005}. A molecular understanding of the variation of this friction coefficient is still lacking and we hope that our results could motivate numerical simulations.

\section{Conclusion}
As a conclusion, we have developed a new experimental setup probing slip at the wall for fluorescent polymer fluids under Couette-Plan flow. The principle of the measurement relies on a photobleached line going through the fluid and subsequent application of the shear. Then the deformations of this photobleached line are analyzed and used as flow tracers in order to measure the slip lengths (at both surfaces) and the shear rate. A major improvement of this new method relies on the detection method as imaging and image analysis techniques can be used leading to a significantly improved accuracy. We have shown that the z-resolved method could be used to obtain a full determination of the velocity profile inside the fluid, with resolution in the direction normal to the solid surfaces. Combining the analysis of the initial intensity profile with the 2D observation of the fluorescent signal from above the sample leads to a full determination of the displacement field along both the shear direction and the normal to the surface. We have demonstrated the potential of these two techniques by studying the flow of a fluorescent entangled polymer melt onto several weakly adsorbing surfaces as a function of shear rate, and demonstrated that these new techniques can achieve slip measurements with a significant improved accuracy compared to the original approach developed~\cite{durliat_influence_1997}. We have shown that a regime of ideal slip where the melt chains are disentangled from the adsorbed chains and the slip is independent of the shear rate can be reached. In this regime the chemical nature of the surface appears to have a strong effect on the magnitude of the slip. As a consequence, we believe that the proposed technique will help getting a clearer description of the friction mechanisms involved at polymer interfaces where the questions of slip of polymer solutions or the role of disentanglements close to the surface are of major interest in many application going from biolubrication to enhanced oil recovery.

\section{Acknowledgments}
This  work  was  supported  by  ANR-ENCORE program (ANR-15-CE06-005). We thank F. Boulogne for its technical help.

\bibliographystyle{unsrt}
\bibliography{biblio}

\end{document}